\documentclass[runningheads]{llncs}
\usepackage{comment}
\usepackage{graphicx}
\usepackage{multirow}
\usepackage[table]{xcolor} 
\usepackage[numbers,sort&compress]{natbib} 
\usepackage{booktabs}
\usepackage{rotating}
\usepackage{hyperref}
\usepackage{array}
\usepackage{tablefootnote}
\usepackage{amsmath}
\usepackage[htt]{hyphenat}

\begin{document}
%
\title{\texorpdfstring {Improving cross-domain brain tissue segmentation in fetal MRI with synthetic data\thanks{This research was funded by the Swiss National Science Foundation (215641), ERA-NET Neuron MULTI-FACT project (SNSF 31NE30\_203977), UKRI FLF (MR/T018119/1) and DFG Heisenberg funding (502024488); We acknowledge the Leenaards and Jeantet Foundations as well as CIBM Center for Biomedical Imaging, a Swiss research center of excellence founded and supported by CHUV, UNIL, EPFL, UNIGE and HUG.}}{Improving cross-domain brain tissue segmentation in fetal MRI with synthetic data}}
\titlerunning{Cross-domain brain tissue segmentation with synthetic data}
%
\author{Vladyslav Zalevskyi\inst{1,2}
\and Thomas Sanchez\inst{1, 2}
\and Margaux Roulet \inst{1, 2}
\and Jordina Aviles Verddera  \inst{3}
\and Jana Hutter   \inst{3, 4}
\and Hamza Kebiri   \inst{1, 2}
\and Meritxell Bach Cuadra   \inst{2, 1}\\
\email{vladyslav.zalevskyi@unil.ch}
}

\authorrunning{V. Zalevskyi et al.}
%
\institute{Department of Radiology, Lausanne University Hospital and University of Lausanne (UNIL), Lausanne, Switzerland \and CIBM Center for Biomedical Imaging, Switzerland \and Department for Early Life Imaging, School of Biomedical Engineering \& Imaging Sciences, King’s College London, London, UK \and Smart Imaging Lab, Diagnostic Radiology, FAU Erlangen-Nuremberg, Erlangen, Germany}
\maketitle              
\begin{abstract}
Segmentation of fetal brain tissue from magnetic resonance imaging (MRI) plays a crucial role in the study of \textit{in utero} neurodevelopment. However, automated tools face substantial domain shift challenges as they must be robust to highly heterogeneous clinical data, often limited in numbers and lacking annotations.
Indeed, high variability of the fetal brain morphology, MRI acquisition parameters, and super-resolution reconstruction (SR) algorithms adversely affect the model's performance when evaluated out-of-domain.
In this work, we introduce FetalSynthSeg, a domain randomization method to segment fetal brain MRI, inspired by SynthSeg. Our results show that models trained solely on synthetic data outperform models trained on real data in out-of-domain settings, validated on a 120-subject cross-domain dataset. Furthermore, we extend our evaluation to 40 subjects acquired using low-field (0.55T) MRI and reconstructed with novel SR models, showcasing robustness across different magnetic field strengths and SR algorithms. Leveraging a generative synthetic approach, we tackle the domain shift problem in fetal brain MRI and offer compelling prospects for applications in fields with limited and highly heterogeneous data.
\end{abstract}
\section{Introduction}
\begin{figure}[!t]
    \centering
    \includegraphics[width=1\linewidth]{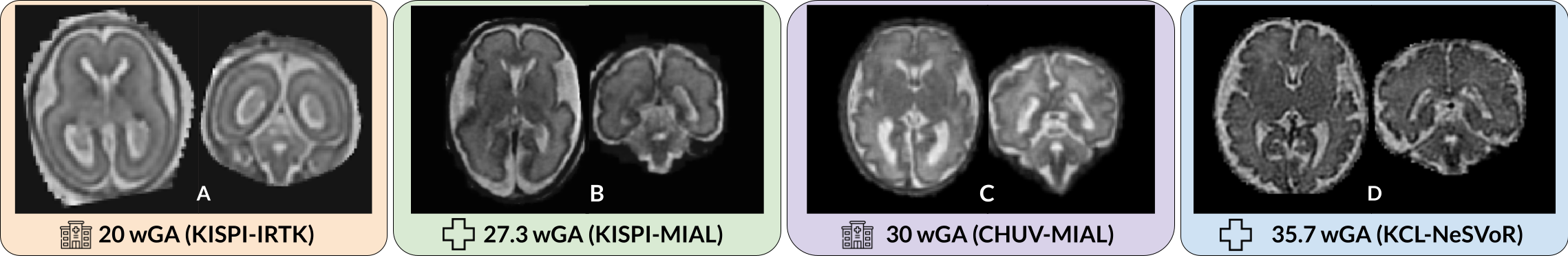}
    \caption{Domain shifts across data splits in fetal SR MRI. (GA in weeks, site-SR). A \& C - pathological, B \& D - neurotypical.}
    \label{fig:ds_examples}
\end{figure}

\looseness=-1
Fetal brain magnetic resonance imaging (MRI) is a growing diagnostic tool for studying neurodevelopment in fetuses~\cite{Benkarim2017,Jakab2021,AvilesVerdera2023,cortes2024mr}. Despite its potential, creating automated pipelines for fetal MRI faces challenges due to the limited availability of annotated datasets and data heterogeneity. The fetal brain undergoes significant morphological changes during gestation and can be severely altered by pathologies, which complicates its automatic analysis~\cite{ch_st_fetatlas2021, Pfeifer2019, diagnostics13142355}. Additionally, datasets encounter distribution shifts from variations in acquisition sites, scanners, and imaging protocols~\cite{Guan2022,lin2023cross}. Utilizing super-resolution-reconstructed (SR) volumes addresses issues like fetal motion artefacts~\cite{sanchez2023fetmrqc} and low through-plane acquisition resolution, but introduces additional heterogeneity in texture, tissue contrast, intensity values and other reconstruction artefacts~\cite{xu2023nesvor} (see Figure~\ref{fig:ds_examples}).
A recent study \cite{payette2024multi} on the FeTA 2022 MICCAI challenge revealed significant performance drops in white matter (WM), gray matter (GM), and ventricles segmentation when models were tested on diverse clinical datasets, highlighting the impact of domain shifts on the analysis of fetal MRI.

Numerous techniques exist to mitigate domain shifts, including domain adaptation \cite{de2022synthetic,xu2023asc}, transfer learning \cite{karani2018_trlearn}, meta-learning \cite{Li2018_metalearning}, style transfer \cite{zhou2021domain} and data harmonization~\cite{HU2023120125}. However, these methods typically rely on the availability of at least target domain images which is challenging in domains with limited data. Other approaches that use multi-centre learning~\cite{Li2018_episod, Liu2020_shaweaware} require multiple training domains which are costly and labour-intensive to acquire. In fetal brain imaging, where new SR algorithms and MRI scanners introduce significant diversity ~\cite{marques2019low}, available domains may not sufficiently capture the required variability for generalizable models.
Recent advancements in single-source domain generalization (SSDG) involve techniques such as global intensity non-linear augmentation (GIN) and causal interventions~\cite{Ouyang2023}. These modifications eliminate spurious correlations, enhancing model robustness to variations in image intensities and textures. In a related study~\cite{Heng2023}, authors analyzed frequency's effect on domain discrepancy, using a mixed frequency spectrum for self-supervised augmentation. While effective, limitations may arise in inducing spatial and intensity transformations, especially with diverse SR algorithms introducing artefacts associated with skull stripping or the inclusion of extra-cerebral tissue.

A promising alternative approach involves synthetic data generation based on segmentation maps~\cite{billot2021synthseg, billot2023robust}, achieving domain generalization through domain randomization~\cite{tremblay2018training}, only requiring labels but no images. Leveraging shape information from segmentations, these models introduce diverse spatial and intensity transformations, along with flexible artefact simulations, mitigating many factors causing domain shifts in MRI.

In our study, we delve into exploring how synthetic generative models can be used to construct a diverse fetal brain dataset for training segmentation models. Our contributions are the following: i) We adapt the domain randomization of Billot et al.~\cite{billot2021synthseg} to fetal brain MRI, accommodating specific fetal anatomical properties, acquisition artefacts and heterogeneity due to fetal brain development and SR algorithms; ii) We show that our method, trained only using synthetic data, performs better than models trained using real data when evaluated out-of-domain and performs on par with state-of-the-art SSDG algorithms; iii) We extend our evaluation to low-field (0.55T) MRI data, showing the robustness of our approach to unseen magnetic field strength and SR algorithms.
\section{Methodology}
\looseness=-1
\subsection{Data}
Various datasets are used in our experiments to validate the SSDG efficacy of the models we explore. These datasets come from multiple institutions and were acquired using MRI scanners from various manufacturers, with different field strengths, acquisition parameters and reconstructed with different SR algorithms. The acquisition details are given in Table~\ref{tab:tab1}.

\noindent\textbf{FeTA dataset.}  We used the publicly available data from the MICCAI 2022 FETA challenge (KISPI) \citep{payette_automatic_2021,fetav2_zenodo}. It consists of 80 subjects, among which are 40 reconstructed using MIALSRTK~\cite{tourbier_mialsuperresolutiontoolkit_2020} and 40 using Simple-IRTK \cite{kuklisova-murgasova_reconstruction_2012}. Expert annotators delineated seven tissue labels (external cerebrospinal fluid (eCSF), GM, WM, ventricles, cerebellum, deep GM, and brainstem). The ethical committee of the Canton of Zurich, Switzerland approved the prospective and retrospective studies that collected and analysed KISPI MRI data (Decision numbers: 2017-00885, 2016-01019, 2017-00167).

\noindent\textbf{Clinical 1.5T dataset.} Additionally, we include a proprietary clinical dataset, named \texttt{CHUV}, containing 40 subjects reconstructed with MIALSRTK~\cite{tourbier_mialsuperresolutiontoolkit_2020} and manually annotated following the FeTA protocol~\cite{payette_automatic_2021}. Data was retrospectively collected from acquisitions done between January 2013 to April 2021. All images were anonymized. This dataset is part of a larger research protocol approved by the ethics committee of the Canton de Vaud (decision number CER-VD 2021-00124) for re-use of their data for research purposes and approval for the release of an anonymous dataset for non-medical reproducible research and open science purposes. This private dataset was used to evaluate methods submitted for the FETA 2022 challenge and a detailed description of it can be found in \cite{payette2024multi}.

\noindent\textbf{Clinical 0.55T dataset.} We also evaluate our model on 40 neurotypical cases acquired on a low-field scanner from another centre, referred to as \texttt{KCL}. Subjects were SR reconstructed twice, using two novel methods, NeSVoR \cite{xu2023nesvor} and SVRTK \cite{uus2022automated}. There are no available manual segmentations for this dataset. Fetal MRI was acquired at Kings College London and approved for sharing with interested academic researchers around the world by the Ethics Committee London Bromley (Ethics code 21/LO/0742). The data has been acquired during a prospective single-center study and has been fully anonymised in line with local procedures

\subsection{FetalSynthSeg}
\begin{figure}[ht!]
    \centering
    \includegraphics[width=1\linewidth]{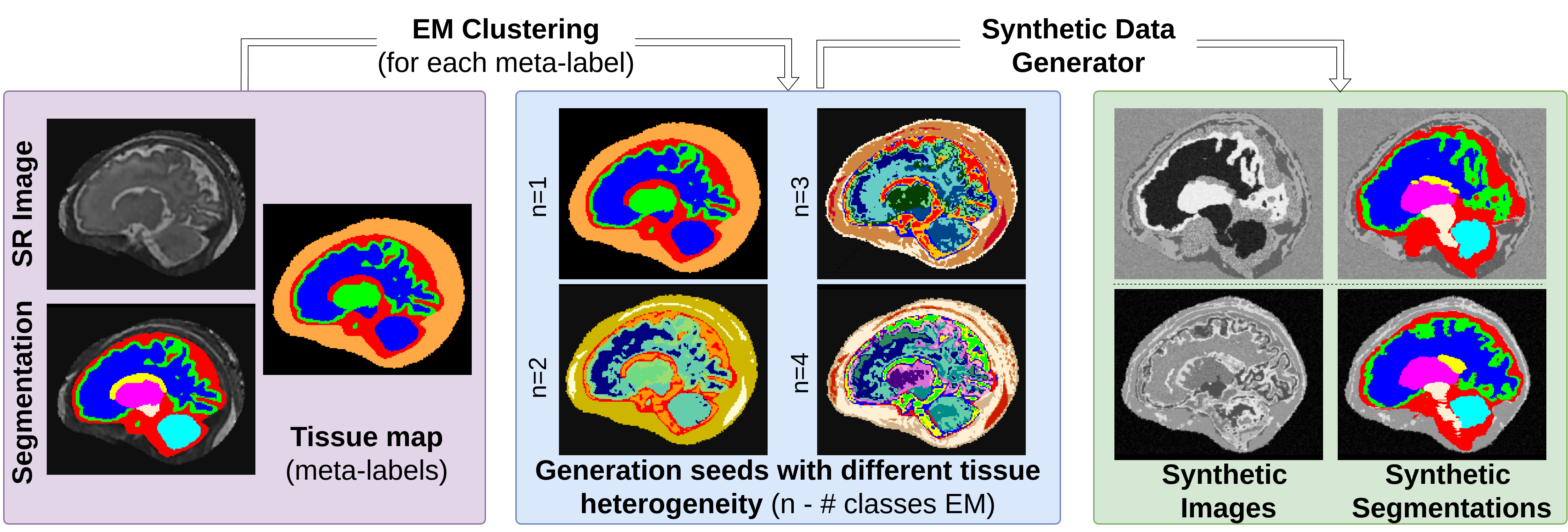}
    \caption{Synthetic image generation framework. Original segmentation labels are merged to create a 4-meta-label tissue map (CSF, WM, GM, skull). EM clustering then divides each meta-label into 1 to 4 subclasses, capturing tissue heterogeneity. A generative model uses these split meta-labels to produce synthetic images.}
    \label{fig:fetSynthGen}
\end{figure}

Our generative model is inspired by SynthSeg~\cite{billot2021synthseg} which is based on domain randomization~\cite{tremblay2018training}. SynthSeg~\cite{billot2021synthseg} leverages image segmentation as a structural prior, integrating randomization across resolution, intensity, contrast, and spatial distortions. This approach yields a diverse dataset that is well-suited for training models capable of robustly handling these sources of variation. We adapt this method to fetal brain segmentation by introducing some crucial changes related to the tissue generation classes (see Figure~\ref{fig:fetSynthGen}).

\begin{table}[ht!]
    \centering
\caption{Dataset properties.}
\centering
\small
\renewcommand{\arraystretch}{1}
\setlength{\tabcolsep}{3pt}
\begin{tabular}{cllccccc}
\hline
    \textbf{Site}                    
    & \textbf{Scanner}                                                                                 
    & \textbf{\begin{tabular}[c]{@{}l@{}}Acquisition \\ Parameters\end{tabular}   }                                           
    & \textbf{SR}     
    & \textbf{\begin{tabular}[c]{@{}c@{}}Res. \\ ($mm^{3}$)\end{tabular}}
    & \textbf{\begin{tabular}[c]{@{}c@{}}GA \\ $(weeks)$\end{tabular}}
    & $N_{n}$             
    & $N_{p}$            
    \\ 
\midrule \addlinespace[1.5mm]
    \multirow{2}{*}{\rotatebox[origin=c]{90}{KISPI*}} 
    & \multirow{2}{*}{\begin{tabular}[c]{@{}l@{}}GE Signa Discovery \\ MR450/MR750 \\ (1.5T/3T)\end{tabular}} 
    & \multirow{2}{*}{\begin{tabular}[c]{@{}l@{}}ssFSE\\ TR: 2500-3500/120 ms\\ 0.5 x 0.5 x 3.5 $mm^3$\end{tabular}} 
    & mial   
    & $0.5^{3}$                               
    & 20-34                                                 
    & 25                  
    & 15                 
    \\ 
\addlinespace[1.5mm] \cline{4-8} \addlinespace[1.5mm]
    &                                                                                         
    &                                                                                                                
    & irtk   
    & $0.5^{3}$                               
    & 20-35                                                 
    & 24                  
    & 16                 
    \\ 
\addlinespace[1.5mm] \hline \addlinespace[0.5mm]
    \rotatebox[origin=c]{90}{CHUV*}                  
    & \begin{tabular}[c]{@{}l@{}}Siemens\\ MAGNETOM \\ Aera (1.5T)\end{tabular}                           
    & \begin{tabular}[c]{@{}l@{}}HASTE\\ TR/TE: 1200/90 ms\\ 1.1 x 1.1 x 3 $mm^3$\end{tabular}                      
    & mial   
    & $1.1^{3}$                               
    & 21-35                                                 
    & 25                  
    & 15                 
    \\ 
\addlinespace[0.5mm] \hline \addlinespace[1.5mm]
    \multirow{2}{*}{\rotatebox[origin=c]{90}{KCL}}  
    & \multirow{2}{*}{\begin{tabular}[c]{@{}l@{}}Siemens\\ MAGNETOM \\ FREE.MAX (0.55T)\end{tabular}}     
    & \multirow{2}{*}{\begin{tabular}[c]{@{}l@{}}HASTE\\ TR/TE: 2500/106 ms\\ 1.5 x 1.5 x 4.5 $mm^3$\end{tabular}}   
    & svrtk  
    & \multirow{2}{*}{$0.8^3$}              
    & \multirow{2}{*}{21-35}                                
    & \multirow{2}{*}{40} 
    & \multirow{2}{*}{0} 
    \\ 
\addlinespace[1.5mm] \cline{4-4} \addlinespace[1.5mm]
    &
    &
    & nesvor 
    &                                                      
    &                                                       
    &                     
    &                    
    \\ 
\addlinespace[2mm] \hline
\multicolumn{2}{l}{*FeTA Data~\cite{payette_automatic_2021}} &
\multicolumn{6}{r}{$N_n$: neurotypical, $N_p$: pathological}
\end{tabular}
\label{tab:tab1}
\end{table}

Instead of directly using target segmentations as generation classes, our approach first introduces an intermediate seed generation step. In it, we initiate synthetic image generation by defining four primary \textit{meta-labels}: CSF, WM, GM, and skull with surrounding tissue. Then we employ the expectation–maximisation (EM) algorithm \cite{em_algo} for intensity clustering within each meta-class, resulting in 1-4 subclasses per meta-class. 
In a second step, these subclasses serve as inputs for synthetic data generators, producing images and segmentations that faithfully reflect the observed heterogeneity in SR and those intrinsic to fetal brains, for example, related to WM maturation \cite{Lajous2022}.
By leveraging tissue-type subclasses rather than original segmentation labels during synthetic image generation, we mitigate reliance on artificial contrast disparities between original labels, a critical consideration given the heterogeneity within a single class. This strategy is similar to the one used in  \cite{billot2021synthseg} for cross-domain cardiac MRI segmentation, which showed the importance of such splitting in representing the target structures with different levels of heterogeneity. Following subclass creation, voxel intensities are independently sampled from Gaussian distributions with randomly selected means and standard deviations. Random artefact corruptions, including bias field simulation, Gaussian blur and noise addition, are applied to introduce common noise and artefacts prevalent in SR images. 
The generative model applies a battery of spatial transformations (affine and elastic) to simulate spatial distortions. Table~\ref{tab:SynthAugmParam} in the Supplementary Material provides all detailed parameters of the generative model. We use offline image generation, creating 200 synthetic images per real image to train the model based on purely synthetic images, which is referred to as  \texttt{FetalSynthSeg} (or \texttt{\_synth} as a suffix) further in the text.
Code and pre-trained models will be released upon acceptance of the paper.
\subsection{Segmentation model}

\noindent\textbf{Architecture.}
Based on Billot et al.~\cite{billot2021synthseg} and Valabregue et al.~\cite{valabregue2023comprehensive}, our study employed a 3D U-Net with five levels, featuring instance normalization, max-pooling, and upsampling operations in the expanding path. Each level includes a 3 × 3 × 3 kernel convolutional layer with LeakyReLU activation, except the final layer using softmax activation. Starting with 32 feature maps, the initial layer doubles and halves after max-pooling and upsampling layers, respectively. Skip connections facilitate information flow between the contracting and expanding paths. We use the same architecture and training hyperparameters across all datasets and splits to ensure comparability.

\noindent\textbf{Pre-processing.} During model training, both real and synthetic images undergo identical pre-processing steps before being fed to the model, including resampling to 0.5 x 0.5 x 0.5 mm\textsuperscript{3} with centre crop and crop-padding to 256x256x256 (when needed), random contrast adjustment via gamma transformation, random affine transformations (scaling, rotation, shearing, and translation), random Gaussian noise  and smoothing with detailed hyperparameters presented in Supplementary Material Table \ref{tab:SynthAugmParam}. Subsequently, image intensities are normalized between 0 and 1 via min-max normalization.

\noindent\textbf{Training.} Models are trained using Adam ($\text{LR}=10^{-3}$) on a combination of Dice and Cross-Entropy losses \cite{valabregue2023comprehensive} with a \texttt{ReduceLROnPlateau} scheduler~\cite{pytorch} ($\text{factor}=0.1$, $\text{patience}=10$) for up to 500 epochs ($\text{batch size}=1$). Training halts on persistent validation dice plateau in the last 10 epochs. Internal validation used 5 randomly selected cases per split, ensuring 35 real and 7,000 synthetic images ($200 \times 35$) for the training of baseline and FetalSynthSeg respectively.

\subsection{Experimental settings}
\looseness=-1
In our experiments, we compare \texttt{FetalSynthSeg} model to i) a baseline model trained on real images and labels (denoted from now on as \texttt{baseline}), as well as to ii) \texttt{fit\_nnUnet}\footnote{The model is available on the FeTA challenge DockerHub page~\cite{payette2024multi}.}, FeTA 2022 challenge winner, an ensemble of nnUnet models trained with all 80 images from KISPI using the GIN SSDG approach~\cite{Ouyang2023}. We aim to show that our model using only synthetically generated images can outperform models trained on real images when tackling out-of-domain generalization and reach comparable performance to SSDG SotA domain generalization approaches.

\noindent\textbf{Experiment 1 - High-field generalization.} We perform the first domain generalization experiment by using the FeTA data as well as the clinical 1.5T dataset. This setting replicates the evaluations carried out in the FeTA challenge 2022~\cite{payette2024multi}.
We consider three data splits, defined by images acquired at specific sites and reconstructed with specific SR, namely \texttt{KISPI - mial}, \texttt{KISPI - IRTK} and \texttt{CHUV - mial} (each containing 35 real images for the baseline training, and 7000 synthetic images for synth training). Two models are trained on each split, one using the original FeTA data and labels, and another using the synthetic data generated from the labels only. This yields a total of six models (denoted as \texttt{<Site>\_<SR>\_base/synth} in figures). The models trained on one (\texttt{Site}, \texttt{SR}) data split are tested on the two other (\texttt{Site}, \texttt{SR}) pairs.
The mean dice score (\texttt{mdsc}) and 95th percentile Hausdorff distance  of each model variant are reported (95th HD in Supplementary Material), and models are compared using the non-parametric Wilcoxon rank-sum test with Bonferroni correction. The p-value for statistical significance is set to 0.05.

\noindent\textbf{Experiment 2 - Low-field generalization.} We assess our trained model's adaptability to new, unseen data with a dataset comprising 40 neurotypical subjects acquired at \texttt{KCL} on a low-field (0.55T) MRI scanner and reconstructed either using SVRTK~\cite{kuklisova-murgasova_reconstruction_2012} or NeSVoR ~\cite{xu2023nesvor} (none of them used in FeTA data).
We select the top-performing model variant trained on original data and the counterpart trained on synthetic data, both derived from the same data split. As no ground truth is available, we evaluate model predictions by comparing tissue volume growth through gestational age (GA) with FeTA reference data. A second-order polynomial fit with confidence interval is evaluated. 
\looseness=-1
\section{Results}
\subsubsection{High-field generalization.} 
\begin{figure}[ht!]
    \centering
    \includegraphics[width=0.8\linewidth]{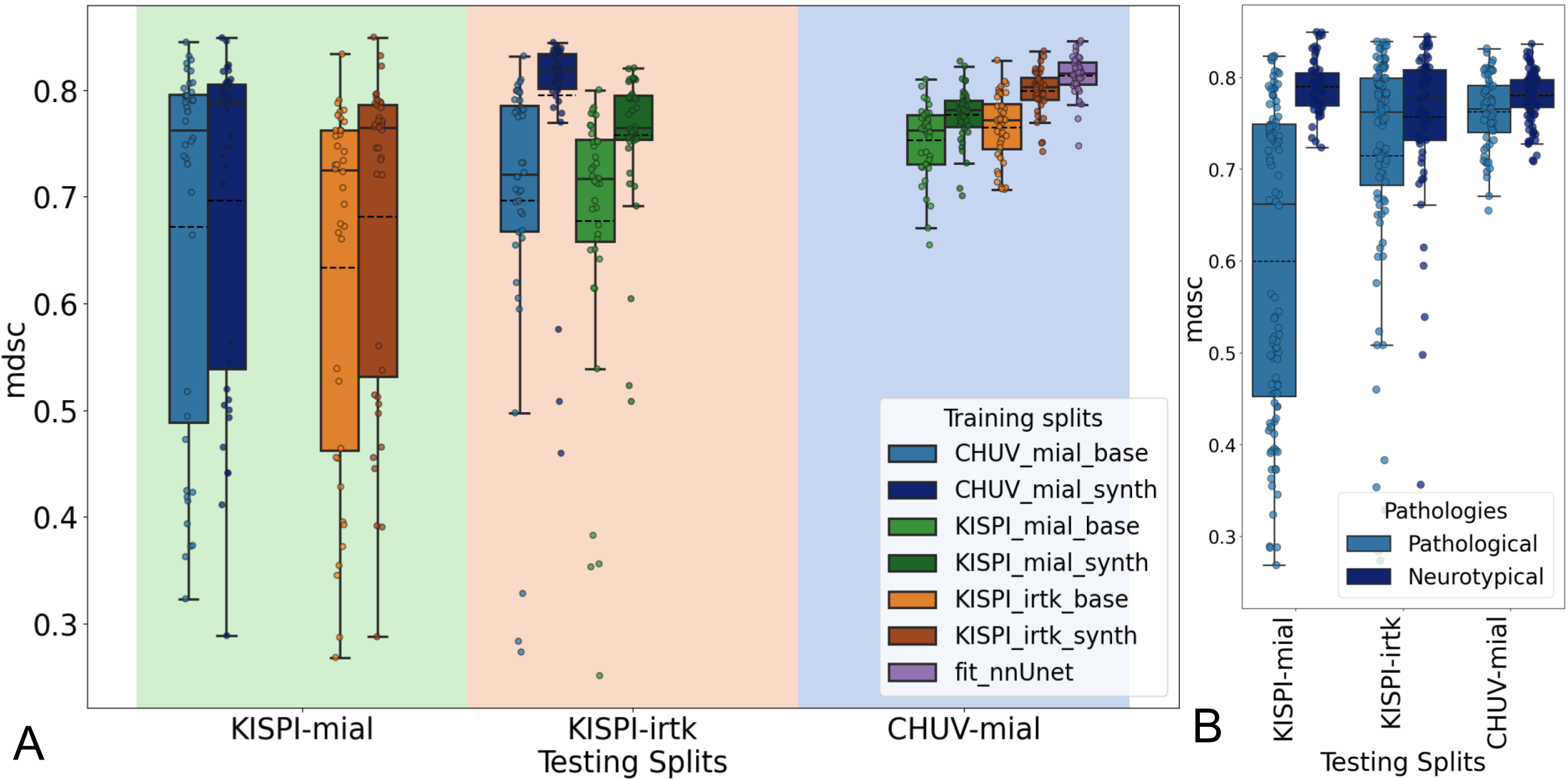}
    \caption{Comparison of the out-of-distribution 
    performance of the segmentation models (mdsc - mean Dice score across all tissues). \textbf{(A)} \texttt{baseline} (light) vs \texttt{FetalSynthSeg} (dark). Data split: \texttt{KISPI-mial} (green), \texttt{KISPI-irtk} (red), \texttt{CHUV-mial} (blue). See Figure \ref{fig:segm_all} from the appendix for a comparison with in-distribution performance as well as results split by gestational age. The dashed horizontal line inside the boxplot corresponds to the mean value. \textbf{(B)} Pathological vs neurotypical, aggregated across all models.}
    \label{fig:synthvsfeta}
\end{figure}

A comparison of the out-of-distribution model predictions is shown in Figure \ref{fig:synthvsfeta}. Models trained on synthetic data consistently outperformed those trained on original images across all out-of-domain testing splits. 
Statistical tests confirmed the significance of the differences in mean Dice scores, with p-values \(<\) 0.00005 for all paired comparisons between corresponding \texttt{baseline} and \texttt{FetalSynthSeg} models trained and tested on the same split. The same trend is observed with the 95th percentile Hausdorff distances, available in Supplementary Figure \ref{fig:segm_all}. 
Additionally, we compared the segmentation accuracy of our model with the FeTA 2022 challenge winners on the \texttt{CHUV-mial} dataset, 
which was not used for \texttt{fit\_nnUnet} training. We highlight that we reached a close albeit slightly lower performance to their solution, although our models were trained on half the amount of data (as they rely on only 35 subjects). 

\begin{figure}[ht!]
    \includegraphics[width=1\linewidth]{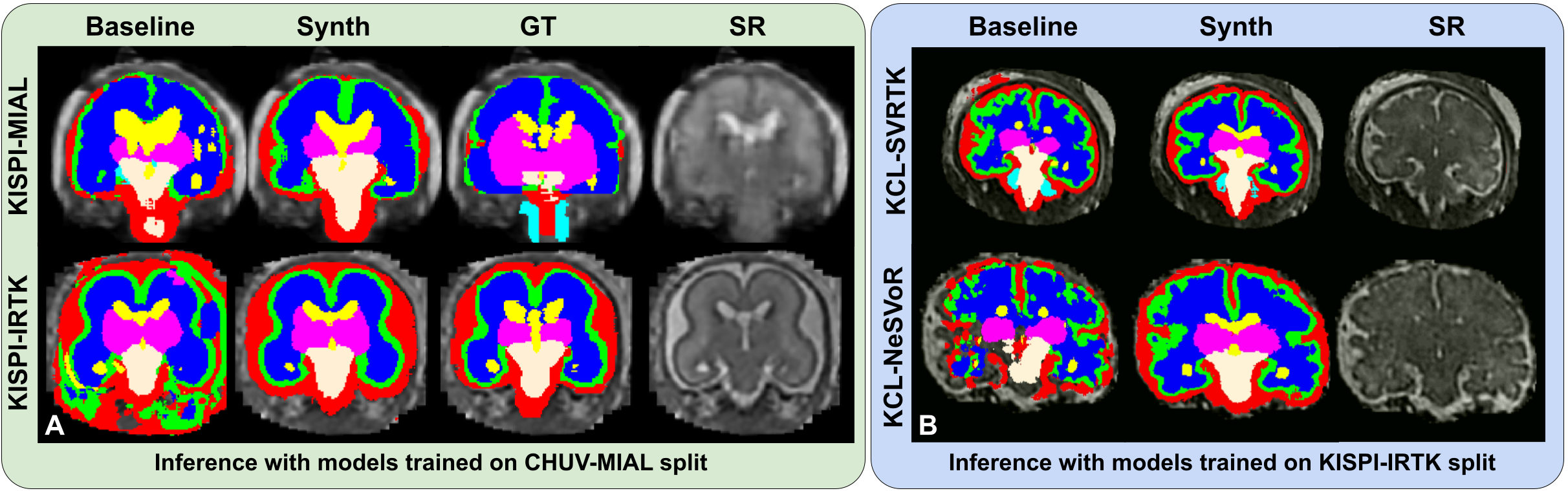}
    \caption{ Cross-domain model inference qualitative results.\textbf{ (A) }Model trained on \texttt{CHUV-mial} and tested on \texttt{KISPI-irtk}/\texttt{KISPI-mial}. \textbf{(B)} Model trained on \texttt{KISPI-irtk} and tested on \texttt{KCL-svrtk}/\texttt{KCL-nesvor}.} 
    \label{fig:synthVSfeta_qual_short}
\end{figure}

The lowest Dice scores were obtained for pathological cases, particularly evident in the \texttt{KISPI-mial} split (see Figure \ref{fig:synthvsfeta}B). This discrepancy can be attributed to the fact that approximately 24\% of the \texttt{KISPI-mial} dataset exhibits poor SR quality often in the severe pathological cases, as noted in \cite{payette_automatic_2021} and illustrated in Figure~\ref{fig:synthVSfeta_qual_short} (top row). 
Nonetheless, we noted qualitative improvements in the \texttt{FetalSynthSeg} model compared to the \texttt{baseline} model, as illustrated in Figure~\ref{fig:synthVSfeta_qual_short} (and in more detail in Figure \ref{fig:FetvsSynthQuallitativeAll} in the Supplement).
Differences in skull stripping by IRTK and MIAL algorithms lead to erroneous segmentation of the skull and surrounding tissue using \texttt{baseline} models. However, the synthetic model is more robust to SR-induced domain shifts and artefacts and avoids these errors, even though it was trained on the segmentations from the same split.
\subsubsection{Low-field generalization.} 
\begin{figure}[ht]
    \centering
    \includegraphics[width=1\linewidth]{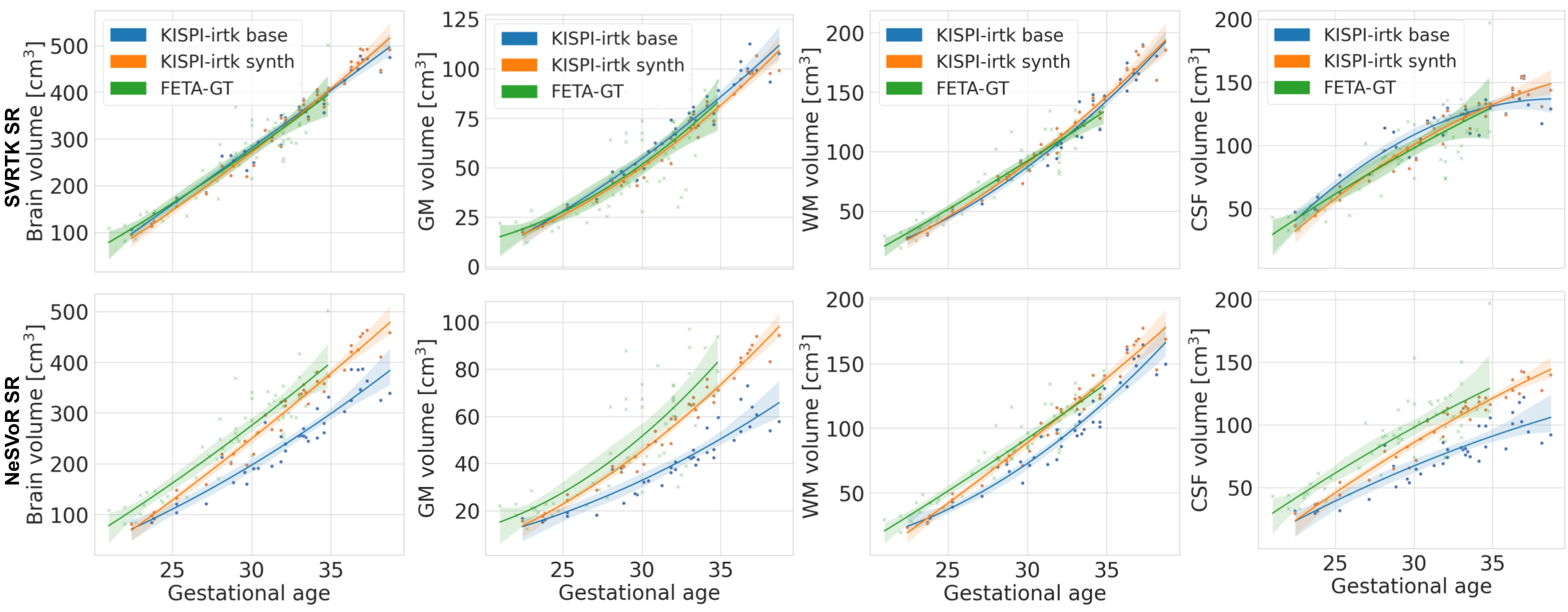}
    \caption{Segmented tissue volumes vs GA for \texttt{KCL} data reconstructed with SVRTK (top row) and NeSVoR  (bottom row). \texttt{KISPI\_irtk\_base} (blue) and \texttt{KISPI\_irtk\_synth} (orange) model predictions are compared to FeTA reference values (green) which are based on the ground truth segmentation of 40 healthy subjects selected across all splits. Lines are second-order polynomial fit and a corresponding shaded area is a confidence interval. See Figure \ref{fig:normALL} in the Supplement for all tissues evaluation.}
    \label{fig:volume_gc}
\end{figure}
Segmented tissue volumes (total brain, GM, WM and CSF volumes) as a function of GA for \texttt{KCL-svrtk} and \texttt{KCL-nesvor} segmentations are illustrated in Figure \ref{fig:volume_gc}. 
The volume growth curves obtained from SVRTK reconstructions remain within the confidence interval for both \texttt{FetalSynthSeg} and \texttt{baseline} models. However, a notable deviation is observed in all estimated tissue volumes for NeSVoR reconstructions predicted by \texttt{KISPI-irtk} \texttt{baseline} model and to a lesser extent from our proposed \texttt{FetalSynthSeg} model. This discrepancy highlights a substantial domain gap within SR algorithms, resulting in an underestimation of expected tissue volumes that do not occur on a closer domain to the KISPI-irtk of SVRTK reconstructions. Remarkably, the model trained on synthetic data demonstrates greater robustness to this domain shift compared to the model trained on real data and exhibits minimal deviation from the expected values calculated on the FeTA dataset while qualitatively showing a superior performance as seen in Figure~\ref{fig:synthVSfeta_qual_short}B.
The baseline model struggles with correct tissue segmentation on NeSVoR reconstructions due to out-of-domain appearance, while performance is improved on SVRTK reconstructions, which exhibit a smaller SR domain gap with IRTK reconstructions. While cortex topology could still be more precise, the synthetic model shows consistent qualitative performance across both scenarios.

\section{Conclusion}

Our study demonstrates that \texttt{FetalSynthSeg} allows robust fetal brain tissue segmentation across datasets with significant domain shifts. We showed how strong randomization of spatial and intensity properties during the synthetic image generation helps models overcome differences caused by MRI acquisition variations and super-resolution reconstruction. Even with half the data, our approach achieved performance close to state-of-the-art SSDG models trained for fetal brain segmentation. 
Compared to models trained solely on real data, those trained exclusively on synthetic data showed superior performance, particularly in cases of novel SR algorithms or images acquired at different field strengths. 
The generalization of segmentation models to low-field MRI is of utmost significance, offering an avenue to enhance fetal MRI accessibility in underserved cohorts and low-income regions, by providing a cost-effective diagnostic solution. 
Our findings suggest that synthetic data can mitigate performance drops caused by limited data and diverse imaging conditions, offering promising applications in fields with highly heterogeneous data, such as fetal imaging.

\bibliography{biblio_v2}
\newpage
\section*{Supplementary material}
\renewcommand\thefigure{S\arabic{figure}}
\renewcommand\thetable{S\arabic{table}}
\setcounter{figure}{0}
\setcounter{table}{0}
\begin{figure}[!ht]
    \centering
    \includegraphics[width=0.75\linewidth]{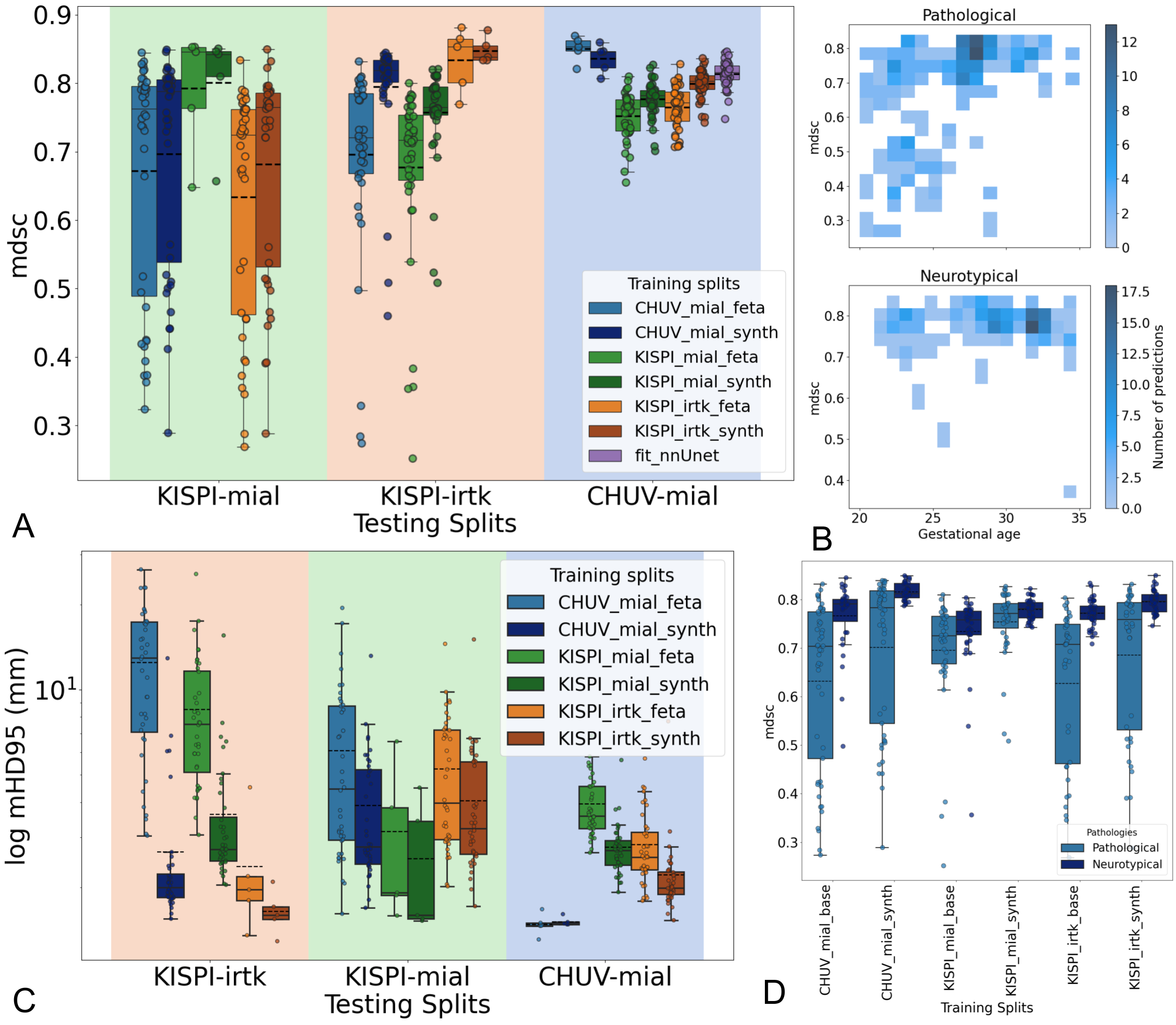}
    \caption{Detailed segmentation evaluation. \textbf{(A)} Comparison between models including in-domain evaluation on the 5 validation cases. \textbf{(B)} Aggregated performance across all models stratified by GA of subjects. \textbf{(C)} Boxplot presenting mean 95th-percentile Hausdorff-Distance scores across all tissues for our experiments on a log scale. \textbf{(D)} Per-model mean dice score results of models evaluated on \texttt{KISPI-mial} split with the distinction between pathological and neurotypical cases.}
    \label{fig:segm_all}
\end{figure}
\begin{table}[hb!]
\centering
\caption{Hyperparameters of the synthetic generator and augmentations. Only parameters deviating from the default ones used by the fully randomized generative model without any tissue priors \cite{billot2021synthseg} are reported. Intensities are in $[0, 255]$ range, rotations in degrees and spatial parameters in mm.}
\small
\begin{tabular}{lclclclc}
\toprule
\multicolumn{8}{l}{Synthetic Data Generator Hyperparameters} \\
\midrule
$a_{\text{sc}}$ & 0.9 & $a_{\text{tr}}$ & -10 & $r_{\text{HR}}$ & 0.5 \\
$b_{\text{sc}}$ & 1.1 & $b_{\text{tr}}$ & 10 & $b_{\text{res}}$ & 0.5 & & \\
\midrule
\multicolumn{8}{l}{Augmentations Hyperparameters\textsubscript{-Probability}} \\
\midrule
$\gamma_{\text{range} - 0.5}$ & 0.5 -- 1.5 & $\text{scale}_{\text{range}- 0.5}$ & -0.1 -- 0.1 & $\sigma_{\text{noise}- 0.5}$ & 0.1 & $\mu_{\text{noise}- 0.5}$ & 0\\
$\text{rotation}_{\text{range}- 0.5}$ & -0.2 -- 0.2 & $\text{shear}_{\text{range}- 0.5}$ & -0.1 -- 0.1 & $\sigma_{\text{smooth}- 0.7}$ & 0.5 -- 1.5 & & \\
\bottomrule
\end{tabular}
\label{tab:SynthAugmParam}
\end{table}
\begin{figure}[ht!]
    \centering
    \includegraphics[width=0.9\linewidth]{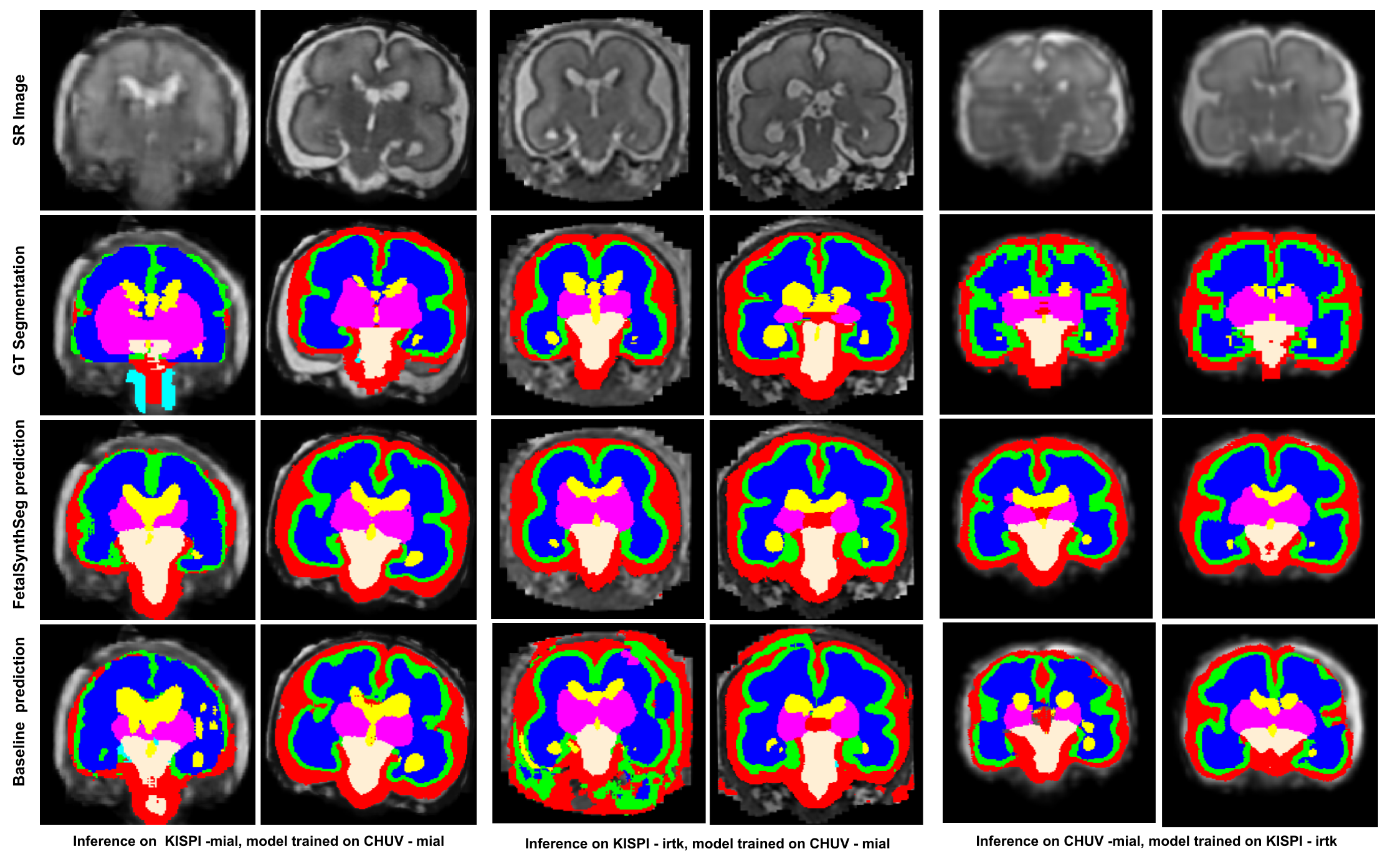}
    \caption{Cross-domain model inference qualitative results}
    \label{fig:FetvsSynthQuallitativeAll}
\end{figure}
\begin{figure}[hb!]
    \centering
    \includegraphics[width=0.93\linewidth]{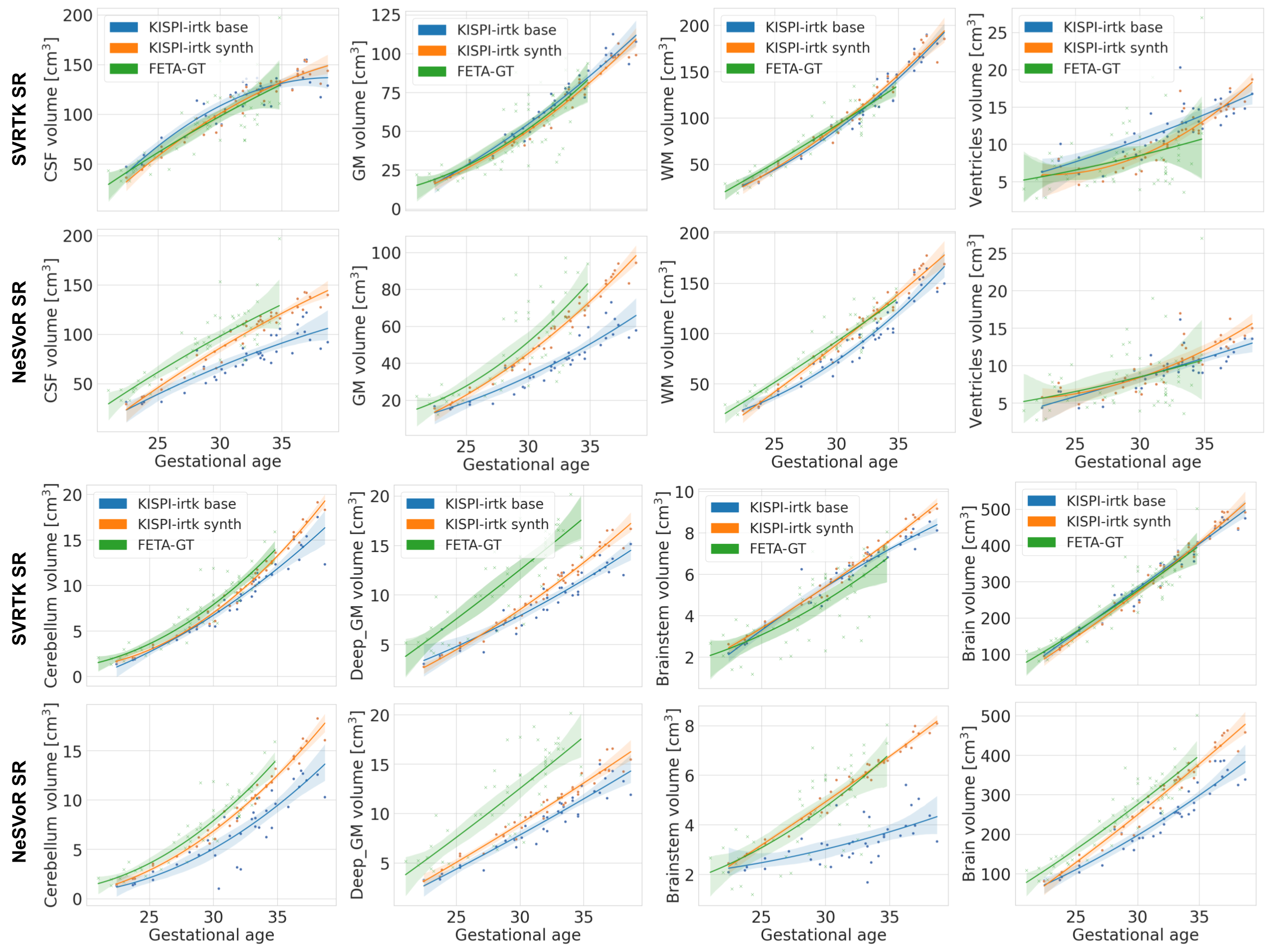}
    \caption{Segmented tissue volumes vs GA for \texttt{}{KCL} data for all tissues.}
    \label{fig:normALL}
\end{figure}

\end{document}